\documentclass[preprint,preprintnumbers,amsmath,amssymb]{revtex4}

\usepackage{graphicx}
\usepackage{dcolumn}
\usepackage{bm}

\begin{document}

\begin{flushright}
YITP-05-46\\
OIQP-05-10
\end{flushright}


\title{Compactified Time and likely Entropy \\ 
-- World Inside Time Machine: Closed Time-like Curve --}

\author{Holger B. NIELSEN}
\affiliation{%
Niels Bohr Institute\\
17, Blegdamsvej, Copenhagen\\
Denmark
}%

\author{Masao NINOMIYA}
 \altaffiliation{Also at
 {Okayama Institute for Quantum Physics, Kyoyama-cho 1-9, 
 Okayama City 700-0015, Japan.}}
\affiliation{
Yukawa Institute for Theoretical Physics,\\
Kyoto University\\
Kyoto 606-8502, Japan
}%

\begin{abstract}
If a macroscopic (random) classical system is put into a random state in phase space, 
it will of course the most likely have an almost maximal entropy
according to second law of thermodynamics.
We will show, however, the following theorem: 
If it is enforced to be periodic with a given period $T$ in advance, 
the distribution of the entropy for the otherwise random state will be much more smoothed out, 
and the entropy could be very likely much smaller than the maximal one. 
Even quantum mechanically we can understand that such a lower than maximal entropy is likely. 
A corollary turns out to be that the entropy in such closed 
time-like loop worlds remain constant.

\end{abstract}

\maketitle

\section{Introduction}
In the present article we shall study a world or a (classical) 
mechanical system 
situated on a closed time-like loop, 
or we can simply consider a model of a universe intrinsically periodic.
This is a study that can be considered a simple exercise playing with ideas 
that could be of relevance for making a (speculative) model 
behind the second law of thermodynamics
\cite{1}\cite{2}\cite{3}\cite{3}\cite{4}.

One problem with combining the second law with time reversal symmetry, 
even if we hope for global features making $\dot{S}\ge 0$ in some era, 
is that then there is  nothing prevent that in an other era 
-- it is at least logically possible -- we then have $\dot{S}\le 0$.
That in turn means that there are some restrictions known not only about the past 
but also about the future.
Such rules about the future will function in a sense as time machine
\cite{5}.

At least we can defend that there is the possibility of ``time machine effects" 
in a model with such regularities about the future 
-- regularities known from a mysterious, presumably not valid law of nature -- 
in the sense that one could
obtain what we should accept as messages from the future.
We could ask:  Why is it 
that we normally do \underline{not} get messages from the future ?

Indeed the point is that a tiny amount of information from or about the past 
can allow us to make great deductions about the past because the past,
which had low entropy, 
is very ordered.
So from a little knowledge we can deduce a lot.
Concerning the future on the other hand we cannot analogously trust the regularities.
So even if we know a tiny amount of photons,
 say to develop into the future 
we cannot count on that they shall be involved with matter.
They can just remain all through the future 
without getting associated with 
any other matters in the future.
We cannot get messages from the future 
because we normally cannot conclude by use of regularity knowledge about the future 
from a tiny amount of information.
For that the future can too easily be irregular.
In our usual world picture (based on second law we might say) 
there is no regularity in the future except for the rudiments 
of regularity
left over from the past.
Therefore we cannot conclude anything great about the future from tiny informations 
like a little bit of light going into future.
Even if we come to know that such a little bit of light runs into the future 
we do not accept that as a message \underline{from the future}, 
because we cannot use that to make further significant deductions about the future.
When we, however, get a letter or a fax, 
sent in the past we can use its content to  
conclude a lot about the past with reasonable reliability.

We want to stress that it is our knowledge about a relatively huge regularity 
in the past that makes it possible for us to consider small letters or light beams 
from the past as messages.

\underline{If} however in some strange violation of the strict second law 
we came to know some regularity about the future, 
then we could begin to accept small pieces of light or paper with text on to give us 
messages about the future.
We would 
then conclude that they would have to go into 
that pattern of regularities, which we had come to know as law of nature.
If we for instance knew that there in some future should exist some very separate 
rather small hot places (time inverted stars) 
that would be  more and more hot by absorbing light, 
then knowing that some light goes out in a certain direction 
will imply such time inverted star(s) in that direction.
Then we could claim it were a message about such a time inverted star.

In any case we argue that because of the conflict 
between time reversal invariance and the second law, 
we can hardly imagine any spontaneous breakdown model 
behind second law providing us with say on era with $\dot{S}\ge 0$ 
unless we also have at least the possibility 
for also messages going the opposite way in time.
So really models that could even have the slightest chance of 
a $T$-invariant model behind second law should have 
at least time machine effect elements in them.

It is therefore reasonable to exercise with as simple 
as possible models with such time machine effect elements in them. 
One of the simplest model of this sort 
for which there can be given several suggestive arguments 
is a world or the mechanical system of the model 
put on a closed time like curve (CTC).
It shall turn out that this model indeed does only a very poor job with respect 
to being a model behind the second law of thermodynamics in 
as far as we shall end up with the conclusion that in such a closed loop time world the entropy 
is almost certainly totally constant, all around the time loop.

Such a time loop model can most simply be taken to mean that 
we postulate the set of moments of time to form an $S^1$-circle 
rather than a straight line 
(or some interval or half time axis in Big Bang theories).
We can simply take it that the general relativity time coordinate 
is of such a nature that going forward in it by a certain constant $T$ (of dimension
of time) 
we come by the coordinate map identifications to
the same time moment 
(or say space-time points) as we started from.
We could call this restriction that the world has got an intrinsic period $T$.

In the usual Big Bang picture which is in simple general relativity enforced 
by the knowledge of the Hubble expansion and suggestive phenomenology 
according to Hawking-Penrose' singularity
 theorem such an intrinsic periodicity 
is of course not possible.

Nevertheless it might be imagined that in a final theory behind general relativity 
and not quite in agreement with the presuppositions of the just mentioned Big Bang 
enforcing theorem, it could e.g. happen that no start or crunch singularities were allowed.

In fact for instance superstring theory is presumably of the type 
without a starting point of time. 
There is possibility that a time parameter would go 
from $-\infty$ to $+\infty$, but it would still mean, in a sense, 
some singularities.
Therefore the most elegant would be a compactified time.

Whatever there might be of essential reasons 
for even nature represented by a compact time or space-time manifold 
it is the purpose of the present article to study a world with closed timelike loops 
-- really with a global forming closed loop -- 
and compactified in space.
As a matter of fact, however, 
we intend to consider so abstract formulation that 
we replace all the fields and their conjugates over the 
whole spatial extension of the compactified universe just by 
one (abstract) phase space.
In other words we do not even consider space explicitly,
 but just a general mechanical system 
 -- i.e. a point moving in phase space according to a Hamilton's equations --. 
The compactness of the time manifold $S^1$ then means 
that this general mechanical system 
is enforced to be strictly periodic with a period $T$ which 
is circumference of the above mentioned $S^1$.

Let us compare what we are concerned with a space-time with holographic principle
\cite{6}.
The holographic principle applies to worlds with a periodicity in time 
-- and thus closed timelike loops -- 
but still having a spatial infinity. 
Then what goes on in this space-time 4-volume 
is then determined by the boundary conditions at the spatial ``far away". 
This is holography 
in the sense that the information on the space-time field configurations 
packed into that on boundary surface placed far out in spatial direction.

In the present article we consider a compactified space 
so that there is no surface far out and then even the state of 
the system should be fixed alone by the structure of the system e.g. 
the Hamiltonian $H$ of the system if we think of it as the abstract mechanical system.

It is a very important point to have in mind for the present article that 
when time is compactified 
-- it means
in reality 
that a periodicity with given period is imposed -- 
there are as many periodicity requirements as the dimension of the phase space.
Also the manifold of initial conditions for the system has the same dimension 
since it simply $is$ the phase space.
Unless some of the periodicity conditions are not independent or perhaps contradictory 
the 
expectation is that there are just a discrete set of solutions to enforced periodicity constraint.
By considering a discrete set of solutions as ``essentially" or ``locally" just one solution 
we can say that the periodicity with enforced 
period $T$ fixes an ``essentially" unique solution.
When the Hamiltonian is 
not explicitly time dependent 
and thus conserved there is formally a \underline{one} dimensional manifold of starting points, 
because a start at one of the points in phase space reached along the orbit of the system 
in phase space will just give rise to the same periodic motion just delayed in its development.
But even that can be called an ``essentially" unique solution.
The main point is that compared to a phase space with a huge number of degrees of freedom 
some discrete solutions or even one dimensional sets of solutions 
is little different from a unique solution.
By far most degrees of freedom get settled by the given period with periodicity fixing.

The main purpose of the present article is to deliver a very general estimate of what 
we would consider the entropy of the essentially unique solution as just stressed above, 
to our periodicity constraint with a fixed period.
To put our estimate in perspective one should have in mind that 
if one chooses a random state of a mechanical system with a probability (density) 
given simply by the phase space measure, one almost certainly find it to have maximal entropy.
The reason that 
it is so is because a macro state is associated with a volume of the phase space proportional 
to $e^{\frac{1}{k} S}$ where $k$ is the Boltzmann's constant and $S$ the entropy.
This is certainly true because the entropy $S$ could be defined to be $k \log $ 
(volume of phase space of macro state)
\cite{7}.

With the above mentioned consideration in mind that thus the normal situation 
for a state chosen randomly in phase space is to have maximal entropy, 
it would be quite interesting to show that 
for a somewhat differently chosen state or point in phase space, 
the entropy is not maximal. 
This is indeed a remarkable thing.

The main point of the present article is precisely to show that 
with high probability the state obtained 
by the enforcement of the given period in advance for a 
``random system" will $not$ have maximal entropy.

In the following section 2 we set up a general formulation for macro states 
and approximately conserved quantities.
In section 3 we argue for our expectations for the likely entropy of the state determined 
from enforced periodicity under some simplified assumptions.
In section 4 we then take into account more general way of effective macro states 
without having them identified completely by sets of exactly conserved quantities.
In section 5 we investigate the extension of our classical calculation of the likely entropy 
to quantum mechanics.
In section 6 we conclude and present some outlook.


\section{The philosophy of random Hamiltonian}
It is the philosophy -- as is usual -- also in the present article to avoid fine tuning
of any parameters,
quantities involved in a given system.
We can strengthen this by making the general assumption that apart from the symmetry restrictions 
which we impose the Hamiltonian and also other parameters or functions describing the model 
are taken to be random.
This kind of randomness means that we e.g. for the Hamiltonian think 
of a probability distribution over the space of a large class of functions defined over phase space.
That is to say that we have in mind a distribution density $P[H]$ destined to multiply 
a functional measure $\mathcal D H$ 
so that the probability for the Hamiltonian belonging to a certain subset ${\mathcal A}$ 
of ``all" functions is given as
\begin{eqnarray}
P_{\mathcal A} = \int _{\mathcal A} P[H]{\mathcal D} H
~.\end{eqnarray}
We may think of it this way, 
but in reality it is very difficult to find a reasonable functional integral measure 
$\mathcal D H$ so that not almost all functions 
after such a measure become 
very bad functions with respect to continuity and differentiability.
For the purposes for which we want such random functions 
it is, however, not so important that the measure be precisely of such a form 
$P[H]\mathcal{D} H$.
Thus we could as $\mathcal D H$ well instead take, 
some system of parameterized functions $H (\boldsymbol{\xi}, \boldsymbol{q}, \boldsymbol{p})$ 
depending on a set of parameters $\xi$.

Then we can choose -- somewhat arbitrarily, 
but still in a reasonable way a measure over the space of parameters.
By making the parameterization so that it guarantees smooth 
(continuously differentiable) 
functions we can in such a way obtain random differentiable functions.
In as far as the Hamilton equations involve the partial derivatives of $H$ 
with respect to the $p_i$'s and the $q_i$'s 
it is of course strictly speaking needed that $H$ be differentiable.
But in practice we care only for properties 
which are true ``almost certainly" in the mathematical sense.
In this way we mainly ignore null-sets 
(i.e. sets of functions with zero measure).
The point of course is mainly that 
we do not accept our conclusions to be proven wrong 
by just a very specially made up counter-example.
We could take the philosophy that 
it would be exceedingly strange if Nature should just have chosen 
a very special function.
We really rather go for investigating what we shall typically expect.
Really we believe that it is in a way a law of nature that 
unless we have some laws enforcing special feature we shall only 
get the most likely results 
according to such a randomness model.
Really of course if we see something unlikely in this randomness philosophy 
we should make a new law explain so that it would no more be strange.
That is how science works, you must make new laws whenever something is strange 
in the sense of violating the old ones or even just the statistical expectations derived 
using only the old laws.

The philosophy described
in this section  is essentially that of random dynamics, 
taking the laws of nature as random.

\section{Setup of formalism for macro states}

The formalism to make definition of entropy possible which we shall use
here is of the following type: We imagine the phase-space of what we
call the macro system divided up to a large number of ``macro states''
characterized by some ``macroscopic variables''. That is to say we
imagine some functions $\vec{\xi}(p,q)$ 
-- called the ``macroscopic
variables'' -- 
to be defined over the phase-space the coordinates of
which we denote symbolically as $(p,q)$. We then think of the set of all
the points in phase space which within some finite small accuracy has
given values $\vec{\xi}_0$  so that
\begin{eqnarray}
 \left|\vec{\xi}_0 - \vec{\xi}(p,q)\right| &<& \varepsilon_{\rm macro},
\end{eqnarray}
as a macro state. 
Here $\varepsilon_{\rm macro}$ denotes a
small quantity for the macroscopic variables. We can then define entropy
$S(\vec{\xi}_0(\varepsilon))$ for the macro state at
$\vec{\xi}(p,q) \approx \vec{\xi}_0$ as 
\begin{eqnarray}
 S(\vec{\xi}_0(\varepsilon))
  &=&
  k \log {\rm Vol}
  \left\{
   (p,q) \left| | \vec{\xi}_0 - \vec{\xi}(p,q) |\right.
   < \varepsilon_{\rm macro}
  \right\}
\end{eqnarray}
where $k$ denotes the Boltzmann constant.

Often in statistical mechanics one may meet macro states. However, the
macro states appeared in the present paper may not quite seem to be of
this kind. Thus it may be needed for us to define that it is in fact
always allowed to work with our point of view.

Examples of the macro states which is not at first glance of the type we
describe is made by almost all Gibbs ensembles and also by grand
canonical ensembles. 
For a simple canonical ensemble one should think of
the temperature as one of the macro state parameters i.e. one of the
$\vec{\xi}_0$-components.  
Formally, all the states in canonical
ensembles corresponding to two different temperatures $T_1$ and $T_2$
say are the same just with different probability (densities). 
Since it
is well known that one can approximate for macroscopic systems
by a
canonical ensemble,
 we could make the temperature of a state in
phase space $(p,q)$ be assigned as a function of the energy $H(p,q)$ for
that state in phase space.
The ``temperature'' $T(p,q)$ of the state
$(p,q)$ should be made
\begin{eqnarray}
 T(p,q) &=& f(H(p,q))
\end{eqnarray}
where f is the function giving the macroscopic relation
\begin{eqnarray}
 T &=& f(U)
\end{eqnarray}
for the system. Here $U$ denotes the energy $U=\left<H\right>$.

Since the spread in energy $H$ for a canonical ensemble becomes rather
small for highly macroscopic system the error by taking the entropy as
we suggested above with $\varepsilon_{\rm macro}$ sufficiently big to
allow the $H$-spread would be small.

It would be easy to argue similarly for grand canonical ensembles by
approximating it by an ensemble with a fixed number of particles. 
Again
we could use the approximately unique relation between the number of
particles and the chemical potential and thus define a macroscopic
variable among the $\vec{\xi}_0$'s to represent the chemical potential.

A priori it may be a question of our possibilities for finding some
parameters which we can keep track of as our ``macroscopic variables''
$\vec{\xi}$.  
Such keeping track of the variables $\vec{\xi}(p,q)$ is
the easiest if they are reasonably stable under the time development
of the system. 
There will be total stability of the macroscopic
variables $\vec{\xi}(p,q)$ 
if they are indeed conserved quantities such
as changes. 
You could typically imagine that they would be taken to be
conserved quantities such as angular momentum, linear momentum etc.

Since the possibility of choosing the macroscopic variables to be
conserved quantities is such a simple way to ensure the stability of
them we shall take this case as to be used in the next section. More
generally, however, one may imagine a situation in which the macroscopic
variables are not totally conserved but rather suffer some diffusion as
time goes on.

\section{The entropy estimate in the simple case of conserved macroscopic
variables}

Let us now illustrate the main argument for that the system with an
enforced periodicity system will be in a state -- or rather go through
a series of states -- in phase space not having maximal entropy. First
we have, however, to specify in what sense we consider random systems or
rather a system with a random dynamics. First of all it means that one
imagines to put some probability measure over the set of all functions H
defined on the phase space. Then one can assume that the Hamiltonian for
the ``random system'' is obtained by picking a random Hamiltonian form H
with a probability distribution given by the above mentioned probability
measure.

To get an estimate of the probability for finding a periodic orbit with
period $T$ 
beginning at $(q_0, \rho_0)$ we may think like the following:
We imagine a system development by the Hamiltonian equations -- using
the random Hamiltonian H -- which was started at $(q_0, \rho_0)$ at,
say, $t=0$. As the Hamiltonian is ``random'' the development in time
will be ``random'' too, except though for the selection rules or
restrictions from that the macro variables are rather stable. In the
simplest case of macro states characterized by (totally) conserved
quantities $I_i$ say the system remains inside the subset of the phase
space corresponding to the starting set of $I_i$-values. The volume in
phase space of the subset to which the system started at $(q_0, \rho_0)$
is 
\begin{eqnarray}
u_p^{2N} \exp(S(\{I_{i0} \})) = u_p^{2N} \exp(S(\{I_i (q_0, p_0) \}))
\label{6}
\end{eqnarray}
where $S(\{I_i (q_0, p_0) \})$ is the entropy of the macro state 
-- characterized by the conserved quantities
$I_i = I_i (q_0, p_0)$ -- corresponding to $(q_0, p_0)$,
and $U_p$ is a factor inserted for each dimension to make the dimensionality correct 
without assigning the entropy very strange dimensionality.
Thinking of defining $S$ by a quantum mechanical formula like 
(\ref{6}) below we should take
\begin{eqnarray}
u_p = \sqrt{h}
\end{eqnarray}
So the smaller this entropy 
$S(q_0,p_0)$ of the starting state $(q_0, p_0)$, so to speak, the smaller 
the volume to which the system of state will be confined.

Now the main thinking is that it is easier or more likely to reach back
to the starting point $(q_0, p_0)$ by accident if the volume into
which the system can move is smaller. Actually the probability to find
back to the starting point $(q_0, p_0)$ just at time $t=T$ (where $T$
is the imposed period) within a given uncertainty range must be
inversely proportional to the phase space volume which is
$e^{S((q_0, p_0))}$.

We could choose a certain fixes accuracy for which in the phase space
by calling that practically we count all $(q,p)$ with
\begin{eqnarray}
 ||(q,p)-(q_0,p_0)||^2 &<& \varepsilon^2_{\rm p.s.}
\end{eqnarray}
as indistinguishable where $\varepsilon^2_{p.s.}$ denotes a chosen small
accuracy error in phase space. For the purpose of judging if the period
is just $T$, we accept
$||(q_0,p_0)-(q(T),p(T))||^2 < \varepsilon_{\rm p.s.}$
as the criterion for this periodicity condition.  Here $(q(T),p(T))$
the time developed ones from $(q_0,p_0)$ during the time $T$. We may
now simply estimate that the probability that $(q_0,p_0)$ with
sufficient accuracy given by $\varepsilon_{\rm p.s.}$ gives rise to a
periodic motion with period $T$ is
\begin{eqnarray}
  &&
  P\left((q_0,p_0) \mbox{ has } T\right) \nonumber\\
  &=&
  P\left[
   (q(T),p(T)),
   \mbox{ is inside $\varepsilon_{p.s.}$ sphere around }
   (q_0,p_0)
  \right] dq_0 dp_0 \nonumber\\
  &\simeq&
  C e^{-S(q_0,p_0)} dq_0 dp_0.
\end{eqnarray}
Here $dp_0 dq_0$ is the (Liouville) phase space measure which is
invariant under canonical transformations. The probability for finding a
period $T$ starting point, i.e. a $(q_0,p_0)$ in a macro state (subset
of phase space) with entropy $S({\rm macro}_1)(=S(q_0,p_0))$ is given
by
\begin{eqnarray}
 C \int_{{\rm macro}_1}e^{-S(q_0,p_0)}dq_0dp_0 
 &=& C e^{S({\rm macro}_1)} \times e^{-S({\rm macro}_1)} \nonumber\\
 &=& C.
\end{eqnarray}
Thus all the possible macro states have the same probability for having a
periodic orbit with period $T$.

This result is remarkable because the different macro states will
typically have wildly different phase space volumes in as far as they
are proportional to $e^{S({\rm macro})}$.

\subsection{How is the realistic situation with respect to $T$-periodic
 orbits?}
 
 In this subsection we should estimate more realistically what the
number of periodic orbits. 
In a macro state with $k$ 
 fixed conserved
quantities in a system with $N$ degrees of freedom so that phase space
has dimension $2N$ this macro state has dimension $2N-k$. By use of
integral invariants of Poincar\'{e}
for small areas extending partly in the direction of a conserved
quantity partly along its conjugate, 
we argue that small line pieces
along these conjugate directions do neither extend nor contract as time
goes on.
This makes $k$ directions of zero scaling. 
The remaining $2N-2k$
dimension will typically scale up or down. 
Let us assume -- we may still have
to prove it -- that half of them i.e. $N-k$ dimensions scale up while
the rest of $N-k$ scales down. 
After long time a little region around
$(q_0,p_0)$ will have scaled up exponentially with ``Lyapunov exponents''
(not only the maximal one, but all positive ones) and will have become
enormous. 
The region obtained from the starting $2N-k$ very small region
will be blown up to a huge $N-k$ dimensional surface, 
still with a bit
of extension in the $k$ directions but strongly contracted in the $N-k$
directions of contraction.

We may imagine reaching such an approximately $N-k$ surface by
propagating the very little region forward from $t=0$ to
$t=\frac{T}{2}$. 
Similarly we should get an effectively $N-k$ dimensional one by
propagating backward from $t=0$ to $t=-\frac{T}{2}$.

To seek periodic solutions for the system we should seek cuts of the two
$N-k$ dimensional surfaces corresponding to $\frac{T}{2}$ and
$-\frac{T}{2}$ respectively both surfaces lying in the $2N-k$
dimensional submanifold in phase space that represents the macro states
characterized by the $k$ fixed conserved quantities ${I_i}$. they will
generically never cut for given ${I_i}$'s because even to just cut in
one point would require generically that the sum of the dimensions of
the surfaces would become equal to the total dimension $2N-k$. Now
$(N-k)+(N-k)= 2N-2k$ and thus the $k$ parameters would have to be tuned
to get a cutting. But $k$ is just the number of the dimensions of the
space of the continuously many macro states and so we expect to find
cutting for discrete values of the macro state characterizing conserved
quantities ${I_i}$.

For the purpose of making an estimate of the number of solutions we
should like to define an effective average of the Lyapunov exponent
$\gamma_{av}$ so as to give the $(N-k)$ dimensional ``area'' of the
$\frac{T}{2}$ surface defined above as 
\begin{eqnarray}
 &
 (\varepsilon_{\rm p.s.})^{N-k}
 \times
 \exp\left[\frac{T}{2} \cdot \gamma_{\rm av} \cdot (N-k)\right]
 &
\end{eqnarray}
Here of course $\gamma_{\rm av} \leq \gamma$ where $\gamma$ is the
maximal Lyapunov exponent. 
We must imagine, if we can consider in
practice the macro state as compact, that the $N-k$ dimensional surfaces
corresponding to $\frac{T}{2}$ and $-\frac{T}{2}$ are much folded back
and forth and essentially cover the whole macro state. 
What we really want to estimate is the typical distances measured 
in the space of the different macro states i.e. in the space with the conserved 
variables $\{I_i\}$ as coordinates.
We ask so to speak how large a $\Delta$ will give us just about one solution 
in the volume $\Delta^k$ in this space with extension $\Delta$.

If we call the distance we imagine to get out in the $\{I_i\}$ from the starting point 
$\Delta$ we cover a space of volume $\Delta^k$.
After such extension we should get generic crossing and could ask how many crossings.

Let us define a ``potential crossing volume" of dimension $2N-k$ 
and imbedded in the macro states as explained of this dimension 
as being the product of the three ``areas": 
\begin{eqnarray}
\varepsilon ^{N-k}_{p, s} \exp(\frac{T}{2}\gamma_{av} (N-k)) \cdot 
\varepsilon ^{N-k}_{p, s} \exp(\frac{T}{2}\gamma_{av} (N-k))\cdot \Delta^k 
~.\end{eqnarray}
The number of solution -- in the $\Delta^k$ volume 
-- is this $(2N-k)$-volume divided by the full volume of the phase
\begin{eqnarray}
u^{2N-k} \exp(S) 
~.
\label{v}
\end{eqnarray}

Now above we though still did not sum over all the cells of size 
$\varepsilon ^{2N-k}_{p, s}$ where the motion could have ``started". 
That of course should be in the phase in question having volume(\ref{v}) 
(see page 19).
So there is place for 
\begin{eqnarray}
\#~\mbox{cells} = \frac{u_{p}^{2N-k}}{\varepsilon_{p, s}^{2N-k}} \cdot \exp(S) 
\end{eqnarray}
small cells.

However we do not really use the extension of the small cells 
in the direction of the $k$ dimensions corresponding to the conjugate variables 
to the conserved quantities, so it is better to think of only counting cells 
in the remaining $(2N-k)-k=2N-2k$ dimensions.
We instead use the $\Delta^k$ volume to count for how far 
we should extend in the $\{I_i\}$-space to get the true crossing.
But that then means that we should count the number of cells being for layers 
in the last $k$ dimensions out of the $2N-k$ but only in the $2N-2k$ ones 
corresponding to the dimensions in which we get the huge extension 
in one or the other time directions.
In this point  of view we should rather say that 
the number of cells we must use is
\begin{eqnarray}
\# (\mbox{cells on surfaces}) &=& 
\frac{u_{s}^{2N-2k}}{\varepsilon_{p, s}^{2N-2k}} 
\exp\left(S \cdot  \frac{2N-2k}{2N-k}\right) \\\nonumber
&=& \left(\frac{u_s}{\varepsilon_{p,s}}\right)^{2N-2k}
\exp\left(S \cdot \left(1- \frac{k}{2N-k}\right)\right) \\\nonumber
&\approx & \left(\frac{u_s}{\varepsilon_{p,s}}\right)^{2N-2k}\cdot \exp (S)
~.\end{eqnarray}
Thus the full number of solutions is 
\begin{eqnarray}\frac{
\varepsilon_{p, s}^{N-k} 
\exp\left( \frac{T}{2} \gamma _{av}(N-k) \right)^2 \Delta ^k
}{u_s^{2N-k}\exp(S)}\cdot 
\frac{u_s^{2N-2k}\exp(S)}{\varepsilon_{p, s}^{2N-2k}
}
~~.\end{eqnarray}

In the light of the number of dimensions $k$ 
corresponding to the conserved quantities being tiny compared 
to the total number of degrees of freedom 
$N\gg k$ we do not consider the factor $\varepsilon_{p, s}^{-k}$ as very important.
If therefore looks (at first) that 
the density of classical periodic solutions with just period $T$ 
is $\frac{1}{\Delta^k}$ evaluated so that  
there is just one solution per k-volume of order $\Delta^k$ which means so that 
\begin{eqnarray}
1\approx 
\varepsilon_{p, s}^{-k}\Delta^k \exp\left(T \gamma _{av}(N-k)\right)
~~.\end{eqnarray}
This density thus becomes 
\begin{eqnarray}
\frac{1}{\Delta^k}\approx
\varepsilon_{p, s}^{-k} \exp \left(T \gamma _{av}(N-k)\right)
~~.\end{eqnarray}
At this stage it thus looks as 
if the density were wildly dependent on the average Lyapunov exponent 
through the presumably hugely varying factor
\begin{eqnarray}
\exp \left(T \gamma _{av}(N-k)\right)
~~.\end{eqnarray}

This would indeed be the result if we took it that 
all the classical solutions had the same probability for being realized.
But as we shall argue in the next subsection this is not realistic.
Rather it will turn out that the bigger 
this huge factor is with the smaller weight should we count the solution in question 
so that actually this huge factor $\exp(T \gamma _{av}(N-k))$ 
gets (essentially) canceled 
and we end up with the result  
-- again -- 
that the density of probability in the space of macro states 
realized 
in the intrinsically periodic world is indeed very smooth,
 slowly varying compared to what huge variation that could have been imaged.

\subsection{Weighting of the different classical solutions}
In this subsection we include a correction
in the sense 
that different classical solutions
\cite{8} 
should not a priori be counted as equally likely but that these different tracks rather
obtain a probability weight strongly related to the Lyapunov exponent.

Realistically 
we should always count that there is some uncertainty,
even if small.
If for no other reason, quantum mechanics will provide such a source of uncertainty.
We shall though still postpone quantum mechanics proper to section 6 
and it may be pedagogical here to think of some other source of uncertainty.

Let us here first develop 
-- for self content -- 
the behavior of orbits very close to a given classical solution
by Taylor expanding the Hamiltonian say.
Defining the deviations of the canonical coordinates $q_i$ 
and the conjugate momenta $p_i$ from their values $q_{cl ~i}$, $p_{cl ~i}$ 
along the considered classical motion
\begin{eqnarray}
\Delta q_i &=& q_i - q_{cl ~i} ~,
\\\nonumber
\Delta p_i &=& p_i - p_{cl ~i}
\end{eqnarray}
we easily derive the Hamiltonian for the Hamilton equations for the deviations
to the accuracy of up to second order terms 
in Taylor expansion  
\begin{eqnarray}
\Delta \dot{q}_i 
&=& \dot{q}_i - \dot{q}_{cl ~i} 
\nonumber\\
&=& \sum_{j} \frac{\partial^2 H}{\partial q_j \partial p_i}\Delta q_i
+\sum_{j}\frac{\partial^2 H}{\partial p_j \partial p_i}\Delta p_j
~,\\\nonumber
\Delta \dot{p}_i &=&
-\sum_{j} \frac{\partial^2 H}{\partial q_j \partial q_i}\Delta q_j  
-\sum_{j} \frac{\partial^2 H}{\partial p_j \partial q_i}\Delta p_j ~.
\end{eqnarray}
Equivalently 
in matrix form
they read
\begin{eqnarray}
\left(
  \begin{array}{c}
    \Delta \dot{q}_1    \\
    \Delta \dot{q}_2   \\
    \vdots     \\
    \Delta \dot{q}_N   \\
    \Delta \dot{p}_1   \\
    \Delta \dot{p}_2   \\
    \vdots   \\
    \Delta \dot{p}_N   \\
  \end{array}
\right)
  = \underline{M}
  \left(
  \begin{array}{c}
     \Delta q_1  \\
     \Delta q_2  \\
     \vdots   \\
     \Delta q_N  \\
     \Delta p_1  \\
     \Delta p_2  \\
     \vdots   \\
     \Delta p_N  \\
  \end{array}
\right)
\end{eqnarray}
where $\underline{M}$
 is the $2N\times 2N$ matrix of 
second derivatives of the Hamiltonian
\begin{eqnarray}
\underline{M}=
\left(
  \begin{array}{cc}
  \frac{\partial^2 H}{\partial q_j \partial p_i}     &   
  \frac{\partial^2 H}{\partial p_j \partial p_i} \\ 
 - \frac{\partial^2 H}{\partial q_j \partial q_i}     &  
 - \frac{\partial^2 H}{\partial p_j \partial q_i}  \\
  \end{array}
\right) 
~~.\end{eqnarray}
Here $N$ is the number of degrees of freedom and the four symbols 
$\frac{\partial^2 H}{\partial q_j \partial q_i}$ etc.
stand for the four $N\times N$ submatrices, with $i$ 
enumerating the rows and $j$ the columns.

Introducing the $2N\times 2N$ matrix
\begin{eqnarray}
\underline{J}=
\left(
  \begin{array}{cc}
  \underline{0}      &    \underline{1}\\
  \underline{-1}     &   \underline{0} \\
  \end{array}
\right)
\end{eqnarray}
we note that the time development matrix $\underline{M}$ obeys
\begin{eqnarray}
\underline{J}~ \underline{M}~ \underline{J}=
\left(
  \begin{array}{cc}
  \frac{\partial^2 H}{\partial p_j \partial q_i}     &    
  \frac{\partial^2 H}{\partial q_j \partial q_i}\\
  -\frac{\partial^2 H}{\partial p_j \partial p_i}     &   
  -\frac{\partial^2 H}{\partial q_j \partial p_i} \\
  \end{array}
\right)
=\underline{M}^T
~.\end{eqnarray}

>From this property we immediately obtain that the spectrum 
for frequencies $\omega$  defined from seeking deviations
\begin{eqnarray}
\left(
  \begin{array}{c}
    \Delta q_1    \\
     \vdots     \\
    \Delta q_N   \\
    \Delta p_1   \\
     \vdots   \\
    \Delta p_N   \\
  \end{array}
\right)
  = e^{-i \omega t}
  \left(
  \begin{array}{c}
     \Delta q_{1~0}   \\
      \vdots   \\
     \Delta q_{N~0}  \\
     \Delta p_{1~0}  \\
      \vdots   \\
     \Delta p_{N~0}  \\
  \end{array}
\right)
\end{eqnarray}
with the $\Delta q_{i~0}$ and $\Delta p_{i~0}$'s being constant 
in time is defined as the eigenvalue spectrum for $\omega$ as 
\begin{eqnarray}
-i \omega 
\left(
  \begin{array}{c}
    \Delta q_{1~0}    \\
     \vdots     \\
    \Delta p_{N~0}   \\
  \end{array}
\right)
  = \underline{M}
  \left(
  \begin{array}{c}
     \Delta q_{1~0}   \\
      \vdots   \\
     \Delta p_{N~0}  \\
  \end{array}
\right)
\label{e}
\end{eqnarray}
or from the zeros of the secular equation
\begin{eqnarray}
0=\mathcal{D}et(\underline{M} +i \omega \underline{1})
\label{s}
\end{eqnarray}
from where it is seen to
consist of pairs of opposite eigenvalues.
That is to say if $\omega$ is an eigenvalue then so is also $-\omega$.
In fact equation (\ref{s}) implies
\begin{eqnarray}
0&=&\mathcal{D}et(\underline{M}^T +i \omega \underline{1})
= \mathcal{D}et(\underline{J}~ \underline{M}~ \underline{J} +i 
\omega \underline{1})\\\nonumber
&=& \mathcal{D}et(\underline{-M} +i \omega \underline{1})
\Rightarrow \mathcal{D}et(\underline{M} -i \omega \underline{1})
\end{eqnarray}
and so we see that together with $\omega$ we must also have $-\omega^*$ 
as an eigenfrequency, namely by complex conjugating equation (\ref{e})
which gives that the complex conjugate column 
$
\left(
  \begin{array}{c}
     \Delta q^*_{1~0}   \\
      \vdots   \\
     \Delta p^*_{N~0}  \\
  \end{array}
\right)
$
is an eigenvector corresponding to the frequency  $-\omega^*$.
Provided  
we do not have a whole set of four different related eigenvalues 
$(\omega, -\omega, -\omega^*, \omega^*)$ 
but rather only two related eigenvalues we must have either
\begin{eqnarray}
\label{*}
-\omega^* = \omega 
~~~\mbox{or}~~
-\omega^* = -\omega
~.\end{eqnarray}

The possibility (\ref{*}) only allows that $\omega$ is 
either purely imaginary or purely real.
Both these possibilities are easily seen to be indeed realizable 
respectively the inverted and usual harmonic oscillator used as examples.

In the case of $\omega$  real the paths close 
to a given classical solution circle around the latter, 
much like a harmonic oscillator 
-- or rather several -- in phase space circles around the equilibrium point.
If we however have the purely imaginary $\omega = -\omega^*$ 
then the situation is rather analogous to ``the inverted harmonic oscillator", 
meaning a particles near the top of a hill.
In such an unstable equilibrium situation 
it is well known that the solutions are rather of the form of 
linear combinations of 
exponentially varying solutions in time
of the form $e^{\pm \gamma t}$ 
call 
the coefficients in the exponent,
$\gamma = Im~\omega$,
in time local Lyapunov exponents.
In any case it is very important for the cancellation, 
which we are going to show that averaging over time these eigenvalues
 $\pm \gamma$ are closely related to our averaged Lyapunov exponent $\gamma_{av}$.
In fact we have the relation
\begin{eqnarray}
\int_{\mbox{the period $T$ along the classical path}}
\sum_{\mbox{the different eigenvalues}} 
|\gamma(t)|dt \frac{1}{T} = \gamma_{av}
~.\end{eqnarray}

In principle $\gamma_{av}$ depends on the path.

The physical meaning of these imaginary frequencies $\omega=i\gamma$ 
is that the nearby paths soon move away from the given classical path 
$(q_{i~0}(t), p_{i~0}(t))$ exponentially.
If we thus put some cut off -- some accuracy of measurement say 
-- for how far out we still consider a neighboring path connected with 
the given classical path
still in the neighbourhood,
 this border 
will be crossed every unit of time by the fraction of 
all the surrounding neighbouring paths being 
\begin{eqnarray}
\label{l}
\mbox{``Fraction being lost per unit time"} = \sum_{|\gamma|}|\gamma(t)|
~.\end{eqnarray}
Note that the probability for some random neighboring path 
is all the time given by (\ref{l})
independent of the precise size and shape of the cut off chosen to define 
whether a path is still in the (closed) neighbourhood.
It is only required that the cut off is chosen to give 
a small enough neighborhood that the Taylor expansion 
we used is valid as a good approximation.
That we have this loss rate (\ref{l}) 
independent of the details of the cut off is of course a consequence of the spreading of 
the neighboring track/paths is a pure (exponential) scaling up with time, 
a scale invariant operation (i.e. not involving any unit for say $q_i$).

When we now consider a classical periodic path with intrinsic period $T$, 
we should realistically think of it 
as representing a tiny little neighborhood of accompanying paths.
Now following along the classical path 
we have just calculated the loss rate of these accompanying path. 
That means that whatever neighborhood 
we had chosen to represent the ``accompanying paths" 
we loose per unit time the fraction $\sum_{|\gamma|}|\gamma(t)|$ of them.
So the total fraction of the accompanying tracks 
which survive all the period $T$ through is
\begin{eqnarray}
\mbox{``Surviving fraction"}= \exp\left(
-\int_{\mbox{along the period}} \sum_{|\gamma|}|\gamma(t)| dt\right)
~.
\end{eqnarray}
This is actually the
``average Lyapunov".
\begin{eqnarray}
\mbox{``Surviving fraction"}= \exp\left(
-\gamma_{av}(\mbox{for the cl. solution)}T
\right)
~.\end{eqnarray}
It is obvious that starting with a random start very close to one of the
periodic solutions the chance that 
it will return a period later equally very close is only this 
$``survival fraction"= \exp\left(-\gamma_{av}(\mbox{for the track)}T\right)$.
We should therefore not take all the classical periodic 
solutions with period $T$
 as equally likely, 
but rather we should weight them statistically with this weight factor
$\exp\left(-\gamma_{av}(\mbox{for the solution)}T\right)$.

Interestingly enough this probability weight 
which a priori could have influenced the relative likelihood for 
different macro states, happens to just cancel the  corresponding factor 
$\exp(\gamma_{av}T)$ occurring in the number of solutions.
That is to say: the effect of loosing the accompanying path 
due to the spreading intervals in time  where some period $\omega$ 
are imaginary is just compensated by the way the number of solutions 
is also sensitive to the Lyapunov exponents.

So we really do not need to calculate Lyapunov exponents to estimate the likelihood 
of the various macro states in our intrinsic periodicity model.
The result that all the macro states have similar probability 
for being realized in the model is still true 
even when the Lyapunov exponents were taken into account 
because their effect is remarkably just canceled out!

\section{More general macro states}

If one makes the assumption used in foregoing section that the
macro states are totally specified by a series ${I_i}$ of conserved
quantum numbers $I_i$, then the macro state cannot change with time at
all. Thus entropy variation with time is completely excluded under such
conditions. Thus any hope of deriving the second law of thermodynamics
in such a model would be excluded
alone for the reason that the macro state could not change.

We therefore must make the assumptions a little more liberal so that
the macro states are able to change the one into the other one. 
These
processes of changing macro state should still be somewhat suppressed 
and
unlikely. Nevertheless it should now be possible.

We wish to argue, however, by an argument similar to that of section 3
that it is unlikely that the periodic orbit with given period T will
have its entropy change, but very little.
Hereby we mean that the
macro state -- if it changes at all -- will only change between
macro states with approximately the same entropy.

The argument runs indeed in the following way: we ask for the
probability of a motion of the system starting at a phase space point
$(q_0,p_0)$ with sufficient accuracy periodic with the period $T$. Let
us say that $(q_0,p_0)$ belongs to a macro state $(q_0,p_0)_M$ with
entropy $S((q_0,p_0)_M)$ where M denotes macro states, but that during
the motion with time $t$ the system runs into other macro states
$M_1,M_2$, etc.. The chance for the system to be at time $t=T$ (by
assuming to at $t=0$) back again at $(q_0,p_0)$ is inversely
proportional to $\sum_i e^{S(M_i)} \approx e^{\max S(M_i)}$. That is to
say that it is inversely proportional to the exponentiated entropy for
that of the passed macro states, which has the biggest entropy,
\begin{eqnarray}
 &\max\left\{S(M_i) \;|\; i \mbox{ corresponding the passed } M\right\}&.
\end{eqnarray}
But the phase space for a starting point $(q_0,p_0)$ in a macro state
$M_k$ is $e^{S(M_k)}=e^{S(q_0,p_0)}$. Thus the chance goes as
\begin{eqnarray}
 e^{S(q_0,p_0) - \max \left\{S(M_i)\right\}} &\leq& 1.
 \label{eq16.eq}
\end{eqnarray}

By having the entropy vary during the periodic motion will -- always --
bring down the relative probability, in fact by the factor given by the
left hand side of (\ref{eq16.eq}).
In this way it gets more and more unlikely the
bigger the spread in the entropy during the passage.

Ignoring the small variation in entropy allowed and it is indeed only a
very small amount allowed, we conclude that the entropy will {\it stay
constant}.

In the above statement we argued completely generally and in an abstract
manner 
with the compactified time manifold of the form $S^1$
that
the initial conditions become essentially fixed and that in such a way
that the entropy becomes completely conserved. 
Note that this means that
in such a machinery 
-- which can be said exists inside a time machine, i.e.
closed time like loop -- 
the entropy becomes constant and there would be
{\it no place for a nontrivial entropy increase}. Therefore the second
law of thermodynamics would only be true in the trivial manner that
entropy stands still. A nontrivial increase would {\it not} be allowed.

Logically it thus seems that since we have certainly the nontrivial
increase in nature we would have to claim that the world would not
possibly turn out to be of type of the compactified time as discussed
in this article.

It could be greatly interesting to remark that as superstring theory is said to
be of the type without the singularities initiating or finalizing the
time-axis, then our way of arguing would imply that superstring theory
would not be compatible with second law of thermodynamics.

We must, however, admit the caveat
that our phase space were taken in the discussion as of finite
volume. Thus if the ``system'' were a field theory in an infinite space,
then perhaps we should reconsider the argument.
Arrow of time axis turn out to infinity
could spoil this argument.

\section{Attempt to extend to quantum mechanics}
 
At first one would have thought that the argument in section 3 that
enforcing periodicity with the period $T$ would lead to essentially all
macro states having equal probability, should easily be extended to work
also for quantum mechanical systems. However, strangely enough we shall
see that at first it does not extend to quantum mechanics.

To repeat the same story quantum mechanically,
in the simplest
case
we should  consider a quantum mechanical system with a series of conserved
operators $I_i$ -- among which $I_i$ is the Hamiltonian -- get that the
imposed periodicity $T$ simply comes to require that 
\begin{eqnarray}
 TH &=& TI_1' \ =\  2\pi n,
 \label{eq17.eq}
\end{eqnarray}
and $I_1'$ the eigenvalue of $I_1=H$
with an integer $n$. 
In this way all the states defined by
\begin{eqnarray}
 &
 \left\{
  \left|\psi\right>
  \left| \; \forall i \;
   [\hat I_i \left|\psi\right> = I_i'\left|\psi\right>]
  \right.
  \right\}
 &
 \label{eq18.eq}
\end{eqnarray}
will be unchanged by the time development operator $e^{-iTH}$ for the
period $T$. 
If we now ask for the macro state specified to have the subspace
characterized by a set of eigenvalues ${I_i}$
the allowed set consists of the set of 
macro
states that satisfy eq (\ref{eq17.eq}).
The dimension of this subspace is obviously to be
identified with $e^{S(M)}$ where $S(M)$ is the entropy of the macro state 
$M$. Here the entropy is given by
\begin{eqnarray}
 S(\{I_i\})
 &=&
 \log \dim
 \left\{
  \left|\psi\right>
  \left| \; \forall i \;
  (\hat I_i \left|\psi\right> = I_i'\left|\psi\right>)
  \right.
 \right\}~.
\end{eqnarray}
Taking seriously that Hamiltonian $H=I_1$, say, is among macro state
specifying parameters, we simply get the condition for the periodic
system with period $T$ as
\begin{eqnarray}
 TH' &=& TI_1' \ =\ 2\pi n
\end{eqnarray}
with an integer $n$. 
It means that the time translation operator through 
the micro states in the macro states in question. 
Actually it is obvious
that on the states in eq (\ref{eq18.eq}), the time translation operator
acts as
\begin{eqnarray}
 e^{-iTH}\left|\psi\right> &=& \left|\psi\right>.
\end{eqnarray}
This simple relation means that the macro states with the imposed period
$T'$ are just those with a series of $I_1'$ values of multiple of
$\frac{2\pi}T$. Such macro states be with a smooth distribution in the
space of $\{I_i'\}$ values. In this manner we reached a result similar
to the one in section 3. The entropy can vary a lot as the $\{I_i'\}$
values are varied and thus we still have the surprise that we get smooth
probability distribution in the manifold of macro state in spite of the
strong variation in entropy $S$ and therefore huge variation of the
phase space volume $e^S$.

\section{Conclusion and outlook}

We have studied the consequences of imposing a given period $T$ in
advance on a ``random'' physical system. What we found were that such a
condition is in general only fulfilled for some or few
macro states. 
The remarkable point is that this macro state is not simply
one with maximal entropy. 
On the contrary we rather get for a random
system same probability for the different macro states
to contain a macro state obeying the
periodicity $T$ requirement.

We got this result both quantum mechanically and classically. It should
be stressed that our quantum mechanical result came by using the
Hamiltonian as one of our macroscopic variables. It should be born in
mind that such a world that has an imposed fixed period $T$ is a world
which can be said to exist inside a time machine. Actually it is obvious
that an enforced periodic time variable running on $S^1$ means a time
machine in the sense that one gets back in time by the periodicity which
simply brings a person -- part of the system we consider -- which
lives into time $T$ back to time $0$. As a consequence of the world
considered -- containing in a time machine, it expected to have ``grand
mother paradoxes'' which have to be resolved by some miracles. We can
consider the surprisingly low entropy resulting for a random model of
the type of the fixed period as an expression of the ``miraculous''
solution to the inconsistencies otherwise easily popping up.

An outlook to an application of the present work could be that we
suppose that for some reason or another -- philosophically that there
be no singularities at which time stops or begins, or superstring theory
-- the time manifold is forced to be compact and thus we are forced to a
periodic world development. 
Then this world would have a promising point
concerning the establishment of a ``deeper understanding'' of the second
law of thermodynamics. However, to get a varying entropy with time is
seemingly not coming, so really it was not a successful model to explain
second law of thermodynamics.

We seek to present our failure to obtain in such a model the genuine 
increase of entropy into a no-go theorem developing the rather trivial point 
that second law is strictly speaking in disagreement 
with the time reversal symmetry principle:
The well known arrow of time problem
\cite{9}. 
Could even some time reversal invariant laws allow 
naturally say in time locally in some era, 
an effective second law to be valid.
We hope in near future to see which conditions are enough to make even such models impossible.
 
\begin{acknowledgments}

This work is supported by Grant-in-Aid for Scientific Research on
Priority Areas, Number of Area 763 ``Dynamics of Strings and Fields'',
from the Ministry of Education of Culture, Sports, Science and
Technology, Japan.
One of us (H.B.N) acknowledges 
for the very helpful discussions
participants of this conference.

\end{acknowledgments}



\end{document}